\documentclass[12pt,letterpaper]{article}
\usepackage{amsmath,amssymb,array,calc,rotating,epsfig,psfrag,amscd, cite}

\makeatletter
\renewcommand\section{\@startsection {section}{1}{\z@}%
                                   {-3.5ex \@plus -1ex \@minus -.2ex}
                                   {2.3ex \@plus.2ex}%
                                   {\normalfont\large\bfseries}}
\renewcommand\subsection{\@startsection{subsection}{2}{\z@}%
                                     {-3.25ex\@plus -1ex \@minus -.2ex}%
                                     {1.5ex \@plus .2ex}%
                                     {\normalfont\bfseries}}
\makeatother

\let\non\nonumber

\let\a=\alpha\let\b=\beta

\let\s=\sigma

\newcommand{\bea}{\begin{eqnarray}}
\newcommand{\eea}{\end{eqnarray}}
\newcommand{\be}{\begin{equation}}
\newcommand{\ee}{\end{equation}}

\newcommand{\p}{\partial}

\newcommand{\C}[1]{$(\ref{#1})$}

\def\IZ{\relax\ifmmode\mathchoice
{\hbox{\cmss Z\kern-.4em Z}}{\hbox{\cmss Z\kern-.4em Z}}
{\lower.9pt\hbox{\cmsss Z\kern-.4em Z}} {\lower1.2pt\hbox{\cmsss
Z\kern-.4em Z}}\else{\cmss Z\kern-.4em Z}\fi}
\def\IR{\relax{\rm I\kern-.18em R}}

\def\one{{\hbox{ 1\kern-.8mm l}}}

\newlength{\bredde}
\def\slash#1{\settowidth{\bredde}{$#1$}\ifmmode\,\raisebox{.15ex}{/}
\hspace*{-\bredde} #1\else$\,\raisebox{.15ex}{/}\hspace*{-\bredde}
#1$\fi}

\newsavebox{\zzzbar}
\sbox{\zzzbar}
  {\setlength{\unitlength}{0.9em}
  \begin{picture}(0.6,0.7)
  \thinlines
  \put(0,0){\line(1,0){0.6}}
  \put(0,0.75){\line(1,0){0.575}}
  \multiput(0,0)(0.0125,0.025){30}{\rule{0.3pt}{0.3pt}}
  \multiput(0.2,0)(0.0125,0.025){30}{\rule{0.3pt}{0.3pt}}
  \put(0,0.75){\line(0,-1){0.15}}
  \put(0.015,0.75){\line(0,-1){0.1}}
  \put(0.03,0.75){\line(0,-1){0.075}}
  \put(0.045,0.75){\line(0,-1){0.05}}
  \put(0.05,0.75){\line(0,-1){0.025}}
  \put(0.6,0){\line(0,1){0.15}}
  \put(0.585,0){\line(0,1){0.1}}
  \put(0.57,0){\line(0,1){0.075}}
  \put(0.555,0){\line(0,1){0.05}}
  \put(0.55,0){\line(0,1){0.025}}
  \end{picture}}

\newcommand{\ena}{\end{eqnarray}}
\newcommand{\beqa}{\begin{eqnarray}}
\newcommand{\eeqa}{\end{eqnarray}}

\renewcommand{\b}{\beta}
\newcommand{\g}{\gamma}

\def\a{\alpha}
\def\b{\beta}

\def\g{\gamma}

\def\s{\sigma}

\begin{document}
\begin{titlepage}

\begin{center}



\vskip 2 cm
{\Large \bf Worldsheet (anti)instanton bound states in type II on $T^2$}\\
\vskip 1.25 cm { Anirban Basu\footnote{email address:
    anirbanbasu@hri.res.in} } \\
{\vskip 0.5cm  Harish--Chandra Research Institute, HBNI, Chhatnag Road, Jhusi,\\
Prayagraj 211019, India}

\end{center}

\vskip 2 cm

\begin{abstract}
\baselineskip=18pt

The 1/8 BPS $D^6\mathcal{R}^4$ coupling in type II string theory compactified on $T^2$ receives contributions from worldsheet instantons and anti--instantons wrapping the $T^2$, up to genus three in string perturbation theory. These involve contributions separately from bound states of instantons and anti--instantons, which are qualitatively similar to such contributions to the 1/2 and 1/4 BPS couplings. At genus two, the $D^6\mathcal{R}^4$ coupling also receives contributions from instanton/anti--instanton bound states unlike the 1/2 and 1/4 BPS couplings, which is a consequence of a T--duality invariant eigenvalue equation a term in the coupling satisfies. We solve this eigenvalue equation to obtain the complete structure of the worldsheet (anti)instanton contributions. In the type IIB theory, strong weak coupling duality leads to certain contributions involving bound states of D string (anti)instantons wrapping the $T^2$.

\end{abstract}

\end{titlepage}


\section{Introduction}

Consider type II string theory toroidally compactified on $T^2$. This maximally supersymmetric theory has a U--duality symmetry group $SL(2,\mathbb{Z}) \times SL(3,\mathbb{Z})$. In the type IIB theory, the non--perturbative\footnote{Here perturbative and non--perturbative are with respect to the string coupling.} $SL(2,\mathbb{Z})_{\tau}$ S--duality symmetry which is inherited from ten dimensions is contained in $SL(3,\mathbb{Z})$ of the U--duality group. The perturbative T--duality  symmetry group is
\be SL(2,\mathbb{Z})_T \times SL(2,\mathbb{Z})_U \ee 
where $T$ and $U$ are the complexified Kahler and complex structure moduli of the $T^2$ respectively. While $SL(2,\mathbb{Z})_U$ directly arises as the $SL(2,\mathbb{Z})$ factor in the U--duality group, the $SL(2,\mathbb{Z})_T$ is contained in $SL(3,\mathbb{Z})$. The moduli dependent coefficients of various amplitudes in this theory when expanded around weak string coupling exhibit a rich perturbative as well as non--perturbative structure. 

In the string frame, the perturbative part of the amplitude takes the form    \be \label{pert}\sum_g (e^{-2\phi}V)^{1-g} f_g (T,\overline{T};U,\overline{U})\ee   
where $\phi$ is the dilaton, and $V$ is the volume of $T^2$ in the string frame metric. The Kahler modulus is given by
\be T = T_1 + i T_2 = B_N +iV,\ee
where $B_N$ is the scalar from the NS--NS sector.
In \C{pert}, the term involving $f_g$ is the genus $g$ amplitude, which involves the T--duality invariant string coupling $e^{-2\phi}V$ as the overall factor. Equality of the IIA and IIB theories on compactifying on $T^2$ yields that $f_g(T,\overline{T};U,\overline{U}) = f_g (U,\overline{U};T,\overline{T})$. Note that the perturbative contribution \C{pert} does not involve states from the Ramond sector.      

The non--perturbative contributions arise from D--instantons as well as from $(p,q)$ string instantons\footnote{We follow the convention of denoting the fundamental string as the $(1,0)$ state.} wrapping $T^2$ where $q \neq 0$, and are exponentially suppressed for large $\tau_2$. For the $n$ D--instanton contribution, the exponentially suppressed factor is of the form 
\be e^{2\pi i (n\tau_1 + i \vert n \vert \tau_2)},\ee   
while for the $(p,q)$ string instanton contribution, it is of the form~\cite{Schwarz:1995dk,Kiritsis:1997em} 
\be \label{nonp}e^{2\pi i T_{p,q}},\ee  
where $T_{p,q} = (q B_R + p B_N) + i \vert p-q\tau\vert V$, where $B_R$ is the scalar from the R--R sector. While the instantons carry positive NS (or R) charge, the anti--instantons carry negative charge.

Let us consider the perturbative contributions given by \C{pert}. Though they are perturbative in the string coupling, they can receive contributions which are non--perturbative in $\alpha'$, the inverse string tension. These contributions arise from worldsheet instantons and anti--instantons\footnote{In fact, \C{nonp} also contains such contributions for $p \neq 0$. However, for the sake of simplicity we restrict ourselves to contributions involving no Ramond sector states.} wrapping $T^2$. While it is difficult to calculate these contributions for generic interactions, the BPS interactions are amenable to a detailed analysis. 

First let us consider the $1/2$ BPS $\mathcal{R}^4$ interaction, where only the terms involving $g=0$ and 1 are non--vanishing in \C{pert}. The worldsheet (anti)instanton contributions are given by~\cite{Green:1997as,Kiritsis:1997em,Green:2010wi}
\be \label{r4}f_1 = 2\pi \sum_{n=1}^\infty \frac{\s_1 (n)}{n} \Big(e^{2\pi i n T} + e^{-2\pi i n \overline{T}}\Big),\ee
where we have ignored all other contributions\footnote{Our normalization is such that $f_0 = \zeta(3).$}. 

For the $1/4$ BPS $D^4\mathcal{R}^4$ interaction, where only terms involving $g=0$, 1 and 2 are non--vanishing in \C{pert}, the worldsheet (anti)instanton contributions are given by~\cite{Basu:2007ru,Green:2010wi}
\bea f\label{d4r4}_1 &=& \frac{4}{\pi} E_2 (U,\overline{U})\sum_{n=1}^\infty \frac{\s_3 (n)}{n^2} \Big(1+\frac{1}{2\pi n T_2}\Big)\Big(e^{2\pi i n T} + e^{-2\pi i n \overline{T}}\Big), \non \\ f_2 &=& \frac{4\pi^2}{3}\sum_{n=1}^\infty \frac{\s_3 (n)}{n^2} \Big(1+\frac{1}{2\pi n T_2}\Big)\Big(e^{2\pi i n T} + e^{-2\pi i n \overline{T}}\Big)\eea
where the Eisenstein series $E_2$ is defined by \C{Eisen} and we have ignored all other contributions\footnote{Our normalization is such that $f_0 = \zeta(5).$}.

Thus \C{r4} and \C{d4r4} both involve an infinite sum of worldsheet (anti)instanton contributions. In fact, each term in the sum results from either instantons or from anti--instantons. This feature changes qualitatively when we consider the $1/8$ BPS $D^6\mathcal{R}^4$ interaction which preserves 4 supercharges. This interaction receives contributions from $g=0$, 1, 2 and 3 in \C{pert}. Again, keeping only terms involving the worldsheet (anti)instanton contributions, we have that~\cite{Basu:2007ck,Green:2010wi}\footnote{Our normalization is such that $f_0 = \zeta(3)^2.$}
\bea \label{diff}f_1 &=& \frac{10}{\pi^2}E_3 (U,\overline{U}) \sum_{n=1}^\infty \frac{\s_5 (n)}{n^3}\Big(1+\frac{3}{2\pi n T_2}+\frac{3}{4\pi^2 n^2 T_2^2}\Big)\Big(e^{2\pi i n T} + e^{-2\pi i n \overline{T}}\Big)\non \\ &&+ 2\pi \zeta(3) \sum_{n=1}^\infty \frac{\s_1 (n)}{n} \Big(e^{2\pi i n T} + e^{-2\pi i n \overline{T}}\Big), \non \\ f_2 &=& 2\pi \Big(E_1 (U,\overline{U})+\frac{\pi}{6}\Big)  \sum_{n=1}^\infty \frac{\s_1 (n)}{n} \Big(e^{2\pi i n T} + e^{-2\pi i n \overline{T}}\Big) + F(T,\overline{T}), \non \\ f_3 &=& \frac{\pi^3}{9}\sum_{n=1}^\infty \frac{\s_5 (n)}{n^3} \Big(1+\frac{3}{2\pi n T_2}+\frac{3}{4\pi^2 n^2 T_2^2}\Big)\Big(e^{2\pi i n T} + e^{-2\pi i n \overline{T}}\Big).\eea
In \C{diff}, $F(T,\overline{T})$ satisfies the eigenvalue equation
\be \label{eigen}\Big(\Delta -12\Big)F(T,\overline{T}) = - 6\Big(E_1 (T,\overline{T})\Big)^2,\ee
where
\be \Delta = 4T_2^2 \frac{\p^2 }{\p T\p \overline{T}}\ee
is the $SL(2,\mathbb{Z})_T$ invariant Laplacian. The relevant Eisenstein series that appear in \C{diff} are defined by \C{Eisen} and \C{E1}. The 1/8 BPS couplings have also been analyzed from the worldsheet perspective in~\cite{Green:1999pv,Gomez:2013sla,DHoker:2014oxd,Basu:2015dqa,Pioline:2015nfa}, and from the spacetime point of view in~\cite{Green:2005ba,Basu:2008cf,Bossard:2014lra,Basu:2014hsa,Pioline:2015yea,Wang:2015jna,Bossard:2015uga,Bossard:2015foa,Bossard:2017kfv,Ahlen:2018wng,Dorigoni:2019yoq,Bossard:2020xod}.

Now in \C{diff} all the contributions apart from that involving $F(T,\overline{T})$ are given by an infinite sum of terms involving either worldsheet instantons or anti--instantons. However, while $F(T,\overline{T})$ yields qualitatively similar contributions separately from bound states of instantons or anti--instantons, it receives additional contributions involving bound states of instantons/anti--instantons because of the presence of the source term in the eigenvalue equation \C{eigen}. In this paper, we analyze the content of \C{eigen} in detail to understand all these contributions at a quantitative level.       
 
\section{The analysis of the eigenvalue equation for $F(T,\overline{T})$}

We now analyze the eigenvalue equation \C{eigen} in detail. To start with, we express $F(T,\overline{T})$ as
\be \label{sum}F(T,\overline{T}) = \sum_{n \in \mathbb{Z}} F_n (T_2) e^{2\pi i n T_1}.\ee
This involves an infinite sum over topologically distinct sectors carrying non--trivial NS charge (the $n=0$ sector carries no charge). 

We shall solve \C{eigen} along with specific boundary conditions. For large $T_2$, we have that $F(T,\overline{T}) \sim T_2^2$ simply because this contribution arises at genus two and this is the large volume scaling. For small $T_2$, the large $T_2$ behavior along with $SL(2,\mathbb{Z})_T$ invariance yields that\cite{Green:2014yxa} 
\be \label{small}F_n (T_2) \sim T_2^{-1}\ee
for all $n$. 

\subsection{The mode carrying no NS charge}

To begin with, let us consider the mode $F_0 (T_2)$ in \C{sum} which carries no NS charge. Using \C{eigen} and the large $T_2$ expansion of $E_1$ given in \C{E1}, we first consider the source terms that are power behaved or logarithmic in $T_2$ in the large $T_2$ limit. We refer to this contribution to $F_0 (T_2)$ as $F_{0,1} (T_2)$.  
Thus we have that
\be \Big( T_2^2 \frac{d^2 }{d T_2^2} -12\Big) F_{0,1} (T_2) = -6\Big(\frac{\pi^2}{3}T_2 - \pi {\rm ln}T_2\Big)^2,\ee 
which has the particular solution
\bea \label{p}F_{0,1} (T_2) =  \frac{\pi^2}{720} \Big(65 -20 \pi T_2 + 48 \pi^2 T_2^2\Big)+\pi^2 {\rm ln} T_2 \Big(-\frac{\pi}{3}T_2 +\frac{1}{2} {\rm ln}T_2 - \frac{1}{12}\Big).\eea
The solution to the homogeneous equation is neglected as $T_2^4$ violates the large $T_2$ boundary condition, while $T_2^{-3}$ is neglected as it violates \C{small}. 

The other contribution to $F_0 (T_2)$ that arises from the source terms that are exponentially suppressed in $T_2$ in the large $T_2$ limit, which we refer to as $F_{0,2} (T_2)$, will be considered later.  

\subsection{The modes carrying NS charge}

We now consider the modes in \C{sum} that carry non--vanishing NS charge. We express the mode $F_n (T_2)$ ($n \neq 0$) which carries $n$ units of NS charge as
\be \label{sumtotal} F_n (T_2) = I_n (T_2) + \sum_{n_i \neq 0, n_1 + n_2 =n} I_{n_1,n_2} (T_2),\ee
where $I_n (T_2)$ and $I_{n_1,n_2} (T_2)$ satisfy the differential equations
\bea \label{e1}\Big( T_2^2 \frac{d^2 }{d T_2^2} -12 -4\pi^2 n^2 T_2^2 \Big) I_n (T_2) = -\frac{24\pi^2 \s_1(n)}{\vert n \vert} \Big(\frac{\pi}{3}T_2 - {\rm ln} T_2\Big) e^{-2\pi \vert n \vert T_2}\eea
and
\bea \label{e2}\Big( T_2^2 \frac{d^2 }{d T_2^2} -12 -4\pi^2 (n_1+ n_2)^2 T_2^2 \Big) I_{n_1,n_2} (T_2) = -\frac{24\pi^2 \s_1(n_1)\s_1 (n_2)}{\vert n_1 n_2 \vert} e^{-2\pi (\vert n_1 \vert + \vert n_2 \vert)T_2}\eea
respectively. 

We now solve \C{e1} and \C{e2} with appropriate choice of boundary conditions. For large $T_2$, the solutions $I_n (T_2)$ and $I_{n_1,n_2} (T_2)$ must have a growth no faster than $T_2^2$ for the same reasons as before\footnote{In fact, we shall see that the solutions are exponentially suppressed, hence exhibiting significantly milder behavior.}. For small $T_2$ each mode has singular behavior no worse than $T_2^{-1}$ in order to satisfy \C{small}.  

\subsubsection{The solution for $I_n (T_2)$}

We express
\be I_n (T_2) = I_n^H (T_2) + I_n^P (T_2),\ee
where $I_n^H (T_2)$ is a solution to the homogeneous equation \C{e1}, while $I_n^P (T_2)$ solves the particular equation \C{e1}. 

Now $I_n^H (T_2)$ is given by~\cite{Green:2014yxa}
\be \label{nh} I_n^H (T_2) = b_n \sqrt{T_2} K_{7/2}(2\pi \vert n \vert T_2)\ee
where $b_n$ is an arbitrary constant. We ignore the other solution $\sqrt{T_2} I_{7/2}(2\pi \vert n \vert T_2)$ since it grows exponentially for large $T_2$, violating our boundary condition. 

The particular solution $I_n^P (T_2)$ is given by
\bea \label{np}I_n^P (T_2) &=& -\frac{\s_1 (n) e^{-2\pi \vert n \vert T_2}}{16\pi n^4 T_2^3}\Big[ -12 \Big(2x^2 + 5x +5\Big){\rm ln}(x/2\pi \vert n \vert) \non \\ &&- \frac{4}{\vert n \vert}P(x){\rm ln}x +4\Big(1+\frac{1}{\vert n \vert}\Big)P(-x) e^{2x}{\rm Ei}(-2x) \non \\ &&- \Big(26x^2 + 95x + 215\Big)-\frac{4}{\vert n \vert}\Big(7x^2 + 25 x+55\Big)\Big],\eea
where $x= 2\pi \vert n \vert T_2$, and $P(x)$ is a polynomial in $x$ defined by
\be \label{P} P(x) = x^3 + 6 x^2 + 15 x+ 15.\ee
Also ${\rm Ei}(-x)$ is the exponential integral function defined in \C{def1}.

To determine $b_n$ using the boundary condition at small $T_2$ mentioned above, we expand both $I_n^H (T_2)$ in \C{nh} and $I_n^P (T_2)$ in \C{np} for small $T_2$. For the solution to the homogeneous equation, we have that
\be \label{cancel1}I_n^H (T_2) = b_n \Big[\frac{15}{16\vert n \vert^{7/2}\pi^3 T_2^3}\Big(1- \frac{2}{5} \pi^2 n^2 T_2^2\Big) +O(T_2)\Big].\ee
For the particular solution we get
\be \label{cancel2}I_n^P (T_2) = -\frac{\pi^2 \s_1 (n)}{2\vert n \vert} \Big[\Big(1- \frac{2}{5} \pi^2 n^2 T_2^2\Big)\frac{\Psi (n)}{(2\pi \vert n \vert T_2)^3} + 4 {\rm ln}(2\pi \vert n \vert T_2)\Big]+O(T_2 {\rm ln}T_2)\ee
where we have used \C{series}, and kept all terms that diverge as $T_2 \rightarrow 0$. Here $\Psi (n)$ is given by the expression
\bea \Psi (n) = -215 - \frac{220}{\vert n \vert} + 60 {\rm ln}(2\pi \vert n \vert) + 60 \Big(\g +{\rm ln}2\Big)\Big(1+\frac{1}{\vert n \vert}\Big). \eea
Note that there is no $T_2^{-2}$ term in \C{cancel2}.
Thus the cancellation of the $T_2^{-3}$ in the small $T_2$ expansion gives us that
\be b_n = \frac{\pi^2 \s_1 (n) \Psi_{3,n}}{15\vert n \vert^{1/2}}\ee
yielding the complete solution. In fact the $T_2^{-1}$ term also cancels on adding \C{cancel1} and \C{cancel2}, and hence the only singular term in $I_n (T_2)$ for small $T_2$ is given by  
\be -2\pi^2 \frac{\s_1 (n)}{\vert n \vert} {\rm ln}(2\pi \vert n \vert T_2).\ee
Now for large $T_2$, $I_n (T_2)$ behaves as $e^{-2\pi \vert n \vert T_2}$ with the leading contribution being given by
\be I_n (T_2)= 2\pi^2\frac{\s_1 (n)}{n^2} {\rm ln}(2\pi \vert n \vert T_2) e^{-2\pi \vert n \vert T_2}\ee
where we have used \C{large}. Thus these are contributions from bound states of worldsheet instantons (or anti--instantons) if $n$ is positive (or negative).

\subsubsection{The solution for $I_{n_1,n_2} (T_2)$}

Like before, we express
\be \label{defI} I_{n_1,n_2} (T_2) = I_{n_1,n_2}^H (T_2) + I_{n_1,n_2}^P (T_2),\ee
where $I_{n_1,n_2}^H (T_2)$ is a solution to the homogeneous equation \C{e2}, while $I_{n_1,n_2}^P (T_2)$ solves the particular equation \C{e2}. 

The solution $I_{n_1,n_2}^H (T_2)$ satisfying the large $T_2$ boundary condition is given by
\be \label{h} I_{n_1,n_2}^H (T_2) = c_{n_1,n_2} \sqrt{T_2} K_{7/2}(2\pi \vert n_1 + n_2 \vert T_2)\ee
where $c_{n_1,n_2}$ is an arbitrary constant.

We now consider the particular solution $I_{n_1,n_2}^P (T_2)$. It is convenient to consider the two cases separately:

{\bf(i)} $n_1$ and $n_2$ have same sign (thus $n_1 n_2 > 0$), and 

{\bf(ii)} $n_1$ and $n_2$ have opposite signs (thus $n_1 n_2 < 0$). 

For case {\bf(i)}, we have that
\be \label{s}I_{n_1,n_2}^P (T_2) = -\frac{6\pi^2\s_1 (n_1)\s_1 (n_2)}{n_1 n_2 x^3} e^{-x}(2x^2 + 5x+5), \ee
where $x= 2\pi \vert n_1 + n_2 \vert T_2$. Unlike the other cases, there are no contributions involving the exponential integral function.

To determine $c_{n_1,n_2}$, we demand the cancellation of the $T_2^{-3}$ term in the small $T_2$ expansion of $I_{n_1,n_2} (T_2)$ as discussed earlier, which gives us that
\be c_{n_1,n_2} = 4\pi^2\frac{\vert n_1 + n_2 \vert^{1/2}}{n_1 n_2 } \s_1 (n_1) \s_1 (n_2). \ee
This also cancels the $T_2^{-1}$ term in the small $T_2$ expansion and hence there are no singular terms in $I_{n_1,n_2} (T_2)$ in this limit, as there is no $T_2^{-2}$ term that arises from \C{s}. 

On expanding $I_{n_1,n_2} (T_2)$ for large $T_2$, we see that all the terms are suppressed by a factor of $e^{-2\pi \vert n_1 + n_2 \vert T_2}$. Hence they arise from bound states of worldsheet instantons or anti--instantons depending on whether $n_1$ is positive or negative. In fact, the leading contribution is given by  
\be \frac{2\pi^2\s_1 (n_1) \s_1 (n_2)}{n_1 n_2 }e^{-2\pi \vert n_1 + n_2 \vert T_2} .\ee

Now for case {\bf(ii)}, we have the particular solution
\bea \label{long}I_{n_1,n_2}^P (T_2) &=& \frac{3e^{-2\pi (\vert n_1 \vert +\vert n_2 \vert)T_2}\s_1(n_1)\s_1 (n_2)}{32\pi^2n_1 n_2 \vert n_1 + n_2 \vert^7 T_2^3}\Big[ (\a-\b)\Big(5(\vert n_1 \vert +\vert n_2\vert)R_{13,15} \non \\ &&+ 10 \pi (n_1+n_2)^2 R_{3,5} T_2 + 4\pi^2(n_1+n_2)^2(\vert n_1\vert+\vert n_2\vert)R_{1,5}T_2^2\Big)\non \\ &&-\frac{\a\b}{2\pi} R_{1,5}\Big(P(2\pi \vert n_1+n_2\vert T_2)e^{\a T_2}{\rm Ei}(-\a T_2)\non \\ &&- P(-2\pi \vert n_1 + n_2 \vert T_2)e^{\b T_2} {\rm Ei}(-\b T_2)\Big)\Big],\eea 
where $\a$ $(>0)$ and $\b$ are defined by
\bea \a &=& 2\pi (\vert n_1\vert +\vert n_2\vert - \vert n_1 + n_2 \vert), \non \\ \b &=& 2\pi (\vert n_1\vert +\vert n_2\vert + \vert n_1 + n_2 \vert),\eea
while $R_{a,b}$ is defined by
\be R_{a,b} = a (n_1 + n_2)^2 - b (n_1 - n_2)^2.\ee
Also $P(x)$ is the polynomial defined by \C{P}\footnote{The expression involving the exponential integral function in \C{long} can be written differently using
\bea && P\Big(2\pi \vert n_1+n_2\vert T_2\Big)e^{\a T_2}{\rm Ei}(-\a T_2)- P\Big(-2\pi \vert n_1 + n_2 \vert T_2\Big)e^{\b T_2} {\rm Ei}(-\b T_2) \non \\ && = {\rm sgn}(n_1){\rm sgn}(n_1 + n_2) \Big[P\Big(2\pi (\vert n_1 \vert -\vert n_2\vert) T_2\Big)e^{4\pi \vert n_2\vert T_2}{\rm Ei}(-4\pi \vert n_2\vert T_2)\non \\ && - P\Big(2\pi (\vert n_2\vert -\vert n_1 \vert) T_2\Big)e^{4\pi \vert n_1\vert T_2} {\rm Ei}(-4\pi \vert n_1\vert T_2) \Big],\eea
where the sign function ${\rm sgn}(x)$ is defined as ${\rm sgn} (x) = 1$ if $x >0$, and ${\rm sgn}(x)=-1$ if $x <0$. }.

In order to determine $c_{n_1,n_2}$, we cancel the $T_2^{-3}$ term in the small $T_2$ expansion of $I_{n_1,n_2} (T_2)$ as before. On using \C{series}, the small $T_2$ expansion of \C{long} is given by 
\bea  \label{Sm}I_{n_1,n_2}^P (T_2) = \frac{3\s_1(n_1)\s_1 (n_2)}{64\pi^3n_1 n_2 \vert n_1 + n_2 \vert^7 T_2^3}\Big[1- \frac{2}{5} \pi^2 (n_1+n_2)^2 T_2^2\Big] \Psi (n_1,n_2) + O(T_2^0)\eea
where
\be \Psi  (n_1,n_2) = -15 \a\b R_{1,5}{\rm ln}(\a/\b)+\frac{5}{2} (\a^2-\b^2)R_{13,15} . \ee
We note that the $T_2^{-2}$ term vanishes in \C{Sm}.
Thus we have that
\be \label{c}c_{n_1,n_2} = -\frac{\s_1 (n_1) \s_1 (n_2) \Psi (n_1,n_2)}{20 n_1 n_2 \vert n_1 + n_2 \vert^{7/2}} .\ee
In fact, the $T_2^{-1}$ term also cancels in the small $T_2$ expansion of $I_{n_1,n_2}(T_2)$ and hence there are no singular terms in this expansion. 

Now consider the large $T_2$ expansion of $I_{n_1,n_2}(T_2)$. For fixed $n_1$ and $n_2$, the leading contribution comes from the homogeneous solution and is of the form $e^{-2\pi \vert n_1+n_2\vert T_2}$. Thus the leading contribution is given by
\be -\frac{\s_1 (n_1) \s_1 (n_2)}{40 n_1 n_2 ( n_1 + n_2 )^4} \Psi (n_1,n_2)e^{-2\pi \vert n_1+n_2\vert T_2}.\ee 
The particular solution is exponentially suppressed by an additional factor of $e^{-\a T_2}$, and the leading contribution is given by
\be -\frac{3 \s_1 (n_1) \s_1 (n_2)}{2 n_1^2 n_2^2 T_2^2}e^{-2\pi (\vert n_1 \vert + \vert n_2\vert) T_2}\ee
on using \C{large}.  

Contributions of this kind that are exponentially suppressed at large $T_2$ arise from bound states of worldsheet instantons and anti--instantons. 

Thus the above expressions yield the complete data needed to evaluate \C{sumtotal}. Now in \C{sumtotal}, the contributions arising from $n_1 n_2 < 0$ yield an infinite sum given by
\be 2 \sum_{n_1 \geq n+1} I_{n_1,n-n_1}\ee  
and hence it is worthwhile to check the convergence of this sum. For this, we focus on the large $n_1$ behavior of the various terms while keeping $n$ fixed. The contribution arising from the particular solution \C{long} is exponentially damped in this limit, hence convergence is trivial. To analyze the contributions that arise from the homogeneous solution \C{h} consider the large $n_1$ limit of $c_{n_1,n-n_1}$ in \C{c}.  This is given by  
\be \label{conv}c_{n_1,n-n_1} \rightarrow \frac{4\pi^2 \s_1 (n_1)^2 \vert n \vert^{7/2}}{35\vert n_1 \vert^5}\ee
as several leading contributions cancel.
Using the inequality~\cite{Robin}
\be \s_1 (n) < e^\g n {\rm ln ln} n +\frac{0.6483n}{{\rm ln ln}n}\ee
for $n \geq 3$, it follows that the sum over $n_1$ is convergent\footnote{A related inequality is given by~\cite{Lagarias}
\be \s_1 (n) < H_n + e^{H_n} {\rm ln}H_n, \ee
for $n > 1$, where $H_n$ is the $n$th harmonic number. Using the asymptotic expansion for $H_n$ given by
\be H_n = {\rm ln}n +\g +\frac{1}{2n} - \sum_{m=1}^\infty \frac{B_{2m}}{2m \cdot n^{2m}}\ee
where $B_m$ are the Bernoulli numbers, we again see that the  
sum over $n_1$ in \C{conv} is convergent.}. 

We now consider $F_{0,2} (T_2)$, as mentioned in section 2.1 which we analyze directly using the solution of \C{e2}. This follows because
\be F_{0,2} (T_2) = \sum_{n \in \mathbb{Z}} {\mathcal{F}}_n (T_2),\ee
where 
\be {\mathcal{F}}_n (T_2) = I_{n,\epsilon-n} (T_2)\ee
as $\epsilon \rightarrow 0$, on using \C{defI}. Potentially singular contributions in this limit all cancel leading to a finite answer. This gives us that
\bea \label{F02}F_{0,2} (T_2) = \frac{3}{14\pi T_2^3} \sum_{n=1}^\infty \frac{\s_1 (n)^2}{n^5}+\sum_{n=1}^\infty Q_n (T_2) e^{-4\pi  n  T_2},\eea
where $Q_n (T_2)$ is given by
\bea \label{qn} Q_n (T_2) = -\frac{\s_1(n)^2}{224 n^5 \pi T_2^3} \Big[24(x +1)^2  +x^4 (2-x) + (x^3 -3)^2 + 15 +x^7 e^{x} {\rm Ei}(-x)\Big],\eea
where $x= 4\pi n T_2$. While the first term in \C{F02} arises from \C{h} on using \C{c}, the second term arises from \C{long}.   

Now using the relation
\be \sum_{n=1}^\infty  \frac{\s_p (n) \s_q (n)}{n^r} = \frac{\zeta(r)\zeta(r-p)\zeta(r-q)\zeta(r-p-q)}{\zeta(2r -p-q)},\ee
we see that 
\bea \label{p1}F_{0,2} (T_2) = \frac{\zeta(3)\zeta(5)}{4\pi T_2^3}+\sum_{n=1}^\infty Q_n (T_2) e^{-4\pi  n  T_2}.\eea
Thus from \C{p} and \C{p1}, we have that
\be F_{0} (T_2) = F_{0,1} (T_2) + F_{0,2} (T_2)\ee
in precise agreement with this result obtained using a different method in~\cite{Basu:2007ck}.  

Using \C{large}, we see that $Q_n (T_2)$ is an infinite series in powers of $T_2$ in the large $T_2$ expansion. Since this contribution is weighted by $e^{-4\pi n  T_2}$, it follows that it arises from the bound state of worldsheet instantons/anti--instantons carrying equal and opposite NS charge $n$. In fact in the large $T_2$ expansion of $Q_n (T_2)$ in \C{qn}, on using \C{large} we see that there are several cancellations at leading orders, which yield the leading contribution
\be -\frac{3\s_1 (n)^2}{ n^4 T_2^2} e^{-4\pi n T_2} \ee
to $F_0 (T_2)$ from the instanton/anti--instanton sector with weight $e^{-4\pi n T_2}$.

\section{S--duality and an elementary consequence for $D$ string instanton contributions}

The worldsheet instanton contributions under S--duality get mapped to $D$ string instanton contributions~\cite{Schwarz:1995dk,Witten:1995im}.  Given the exact expressions for the  worldsheet instanton contributions, though it takes work to implement S--duality in order to obtain the complete $D$ string instanton contributions, it is elementary to implement strong weak coupling duality to obtain a part of the $D$ string instanton contributions, which we now illustrate.  

As a simple example, consider the worldsheet instanton contribution to the $\mathcal{R}^4$ coupling given by
\be \label{ws}2\pi \sum_{n=1}^\infty \frac{\s_1 (n)}{n} e^{2\pi i n T} ,\ee
which follows from \C{r4}. In the background where $\tau_1 =0$, strong weak coupling duality yields
\be \label{S}\tau_2 \rightarrow \frac{1}{\tau_2}, \quad V \rightarrow \tau_2 V, \quad B_N \rightarrow B_R, \quad B_R \rightarrow - B_N.\ee
Thus performing the S--duality transformation \C{S} on \C{ws}, we get the $D$ string instanton contribution\footnote{In fact this is the complete answer from the sum over the $(0,n)$ $D$ string instantons which follows from the U--duality invariant expression for the $\mathcal{R}^4$ coupling~\cite{Green:1997as,Kiritsis:1997em}. } 
\be \label{d} 2\pi \sum_{n=1}^\infty \frac{\s_1 (n)}{n} e^{2\pi i n S},\ee
where
\be S = S_1 + i S_2 = B_R + i \tau_2 V.\ee

Similarly for the $D^6\mathcal{R}^4$ coupling the S--duality transformations \C{S} yield partial contributions to the $D$ string instanton contributions using the various expressions for the worldsheet instanton contributions we have analyzed. For example, from \C{qn} we see that the contribution from the bound states of $D$ string instantons/anti--instantons carrying no net RR charge is given by  
\be \sum_{n=1}^\infty \widetilde{Q}_n (S_2) e^{-4\pi n S_2},\ee 
where $\widetilde{Q}_n (S_2)$ is given by
\be \widetilde{Q}_n (S_2) = -\frac{3\s_1 (n)^2}{n^4 S_2^2}\Big[ 1-\frac{4}{y} +\frac{1}{168} \sum_{m=0}^\infty \frac{(-1)^m (m+7)!}{y^{m+2}}\Big],\ee
where $y=4\pi n S_2$ and we have performed a weak coupling (large $\tau_2$) expansion using \C{large}. While the overall $S_2$ dependence must arise from the structure of zero modes in the instanton/anti--instanton background, we see that the infinite sum is an expansion in $y \sim e^{-\phi}$, the open string coupling. Note that performing \C{S} on $I_n^P (T_2)$ in \C{np} yields contributions having factors of ${\rm ln}\tau_2$, which arise from non--local interactions logarithmic in the external momenta in the string frame, on converting to the Einstein frame. This is precisely what is expected from the structure of the U--duality invariant eigenvalue equation that arises for the $D^6\mathcal{R}^4$ coupling~\cite{Basu:2007ck,Green:2010wi}, as the source term contains ${\rm ln}\tau_2$ that arises from the $\mathcal{R}^4$ coupling~\cite{Kiritsis:1997em}.    

\appendix

\section{The $SL(2,\mathbb{Z})$ invariant non--holomorphic Eisenstein series}

The non--holomorphic Eisenstein series $E_s (T,\overline{T})$ is given by the expression 
\bea \label{Eisen}E_s (T,\overline{T}) &=& 2\zeta(2s)T_2^s + 2\sqrt{\pi} T_2^{1-s} \frac{\Gamma(s-1/2)}{\Gamma(s)}\zeta(2s-1) \non \\ &&+ \frac{4\pi^s \sqrt{T_2}}{\Gamma(s)}\sum_{n \neq 0} \frac{\s_{2s-1}(n )}{\vert n\vert^{s-1/2}}K_{s-1/2}(2\pi \vert n\vert T_2) e^{2\pi i nT_1}\eea
on expanding around large $T_2$. Here the divisor function $\s_m (n)$ is defined by
\be \s_m (n) = \sum_{d|n, d >0} d^m,\ee
where the sum is over the positive divisors of $n$. The case $s=1$ has to be regularized and is given by
\bea \label{E1}&&E_1 (T,\overline{T}) = -\pi {\rm ln}\Big(T_2 \vert \eta (T)\vert^4\Big) \non \\ &&= \frac{\pi^2}{3}T_2 - \pi {\rm ln}T_2  +2\pi\sum_{n \neq 0} \frac{\s_{1}(n )}{\vert n\vert}e^{2\pi i (nT_1 +i\vert n \vert T_2)}.\eea

\section{The exponential integral function} 

The exponential integral function ${\rm Ei} (-x)$ is given by the integral representation
\be \label{def1} e^x {\rm Ei} (-x) = -\frac{1}{x} + \int_0^\infty dt\frac{e^{-t}}{(t+x)^2}, \quad x > 0.\ee
Thus we see that $e^x {\rm Ei} (-x)$ is a polynomial in $1/x$ of the form
\be \label{large}e^x {\rm Ei} (-x) = -\frac{1}{x} +\sum_{n=0}^\infty \frac{(-1)^n (n+1)!}{x^{n+2}}\ee
for large $x$. On the other hand, for small $x$, the series expansion is given by
\be \label{series}{\rm Ei} (-x) = \g + {\rm ln}x +\sum_{n=1}^\infty \frac{(-x)^n}{n\cdot n!}, \quad x > 0.\ee


\providecommand{\href}[2]{#2}\begingroup\raggedright\endgroup

\end{document}